RESEARCH ARTICLE　　　　　　　　　　　　　　　　　　　　　　　　　　　　　　　　　　　　OPEN ⬺ ACCESS

# Space weather challenges of the polar cap ionosphere

Jøran Moen[1,*], Kjellmar Oksavik[2], Lucilla Alfonsi[3], Yvonne Daabakk[1], Vineenzo Romano[3], and Luca Spogli[3]

[1] Department of Physics, University of Oslo, P.O. Box 1048 Blindern, NO-0316 Oslo, Norway
　*corresponding author: e-mail: jmoen@fys.uio.no
[2] Department of Physics and Technology, University of Bergen, P.O. Box 7803, NO-5020 Bergen, Norway
[3] Istituto Nazionale di Geofisica e Vulcanologia, Via di Vigna Murata 605, I-00143 Rome, Italy

**ABSTRACT**

This paper presents research on polar cap ionosphere space weather phenomena conducted during the European Cooperation in Science and Technology (COST) action ES0803 from 2008 to 2012. The main part of the work has been directed toward the study of plasma instabilities and scintillations in association with cusp flow channels and polar cap electron density structures/patches, which is considered as critical knowledge in order to develop forecast models for scintillations in the polar cap. We have approached this problem by multi-instrument techniques that comprise the EISCAT Svalbard Radar, SuperDARN radars, in-situ rocket, and GPS scintillation measurements. The Discussion section aims to unify the bits and pieces of highly specialized information from several papers into a generalized picture. The cusp ionosphere appears as a hot region in GPS scintillation climatology maps. Our results are consistent with the existing view that scintillations in the cusp and the polar cap ionosphere are mainly due to multi-scale structures generated by instability processes associated with the cross-polar transport of polar cap patches. We have demonstrated that the SuperDARN convection model can be used to track these patches backward and forward in time. Hence, once a patch has been detected in the cusp inflow region, SuperDARN can be used to forecast its destination in the future. However, the high-density gradient of polar cap patches is not the only prerequisite for high-latitude scintillations. Unprecedented high-resolution rocket measurements reveal that the cusp ionosphere is associated with filamentary precipitation giving rise to kilometer scale gradients onto which the gradient drift instability can operate very efficiently. Cusp ionosphere scintillations also occur during IMF $B_Z$ north conditions, which further substantiates that particle precipitation can play a key role to initialize plasma structuring. Furthermore, the cusp is associated with flow channels and strong flow shears, and we have demonstrated that the Kelvin-Helmholtz instability process may be efficiently driven by reversed flow events.

## 1. Introduction

One of the first known space weather effects were scintillation disturbances on trans-ionospheric radio waves (Hey et al. 1946). Plasma instabilities cause the otherwise relatively uniform plasma to develop inhomogeneous magnetic field-aligned structures over scale sizes from tens of kilometers to meters (Basu et al. 1990), and they are generally most severe at high and low latitudes (e.g., Basu & Basu 1981; Basu et al. 1988a, 1998), but may occur at any place or time during the entire solar cycle (Kintner et al. 2007).

These irregularities may disrupt Very High Frequency (VHF), Ultra High Frequency (UHF), and Global Navigation Satellite Systems (GNSS) at L-band frequencies. At longer wavelengths the same plasma irregularity regions also backscatter radio waves (e.g., HF and VHF). The amplitude and phase scintillations of the radio wave front cause degraded performance for Global Navigation Satellite System (GNSS) receivers and even loss of satellite lock and navigation solution.

Prikryl et al. (2010) reported that cycle slips frequently occur in the cusp region even during solar minimum conditions.

The generalized view is that polar cap scintillations are created in density gradients associated with polar cap patches. Polar cap patches are defined as 100 km scale regions with F-region plasma densities 2–10 times larger than the background density in the polar cap (Crowley 1996; Rodger & Graham 1996; Crowley et al. 2000), and the gradient drift instability is considered as the dominant instability mode (e.g., Ossakow & Chaturvedi 1979; Keskinen & Ossakow 1983; Buchau et al. 1984; Tsunoda 1988; Basu et al. 1990, 1998; Coker et al. 2004; Prikryl et al. 2011a; Basu et al. 1994; Gondarenko & Guzdar 2004a, 2004b). Recent scintillation studies concur with this general view (e.g., Mitchell et al. 2005). Spogli et al. (2009) presented GPS scintillation climatology in the Scandinavian arctic sector. They revealed cusp and auroral oval boundaries as the most active scintillation regions. They pointed out that asymmetry around midnight of the phase scintillation occurrence is in agreement with the polar cap patch distribution found by Moen et al. (2007). Prikryl et al. (2010) found that strong phase scintillations were observed at high latitudes, while the $S_4$ index remained low. They found that the seasonal and hourly dependencies of phase scintillations have the same occurrence pattern as fixed frequency ionosonde F-region echoes and HF radar backscatter, which agrees well with the model dependencies of patch-to-background $N_mF2$ by Sojka et al. (1994). Prikryl et al. (2011a) found evidence that cusp irregularities are driven by polar cap patches and E × B convection dynamics, while nightside auroral scintillations seem to be driven by energetic electron precipitation.

The ultimate space weather product will be physically based radio wave scintillation forecasts. Prikryl et al. (2012) demonstrated a promising attempt of probabilistic forecasting of GPS phase scintillation at high latitudes due to high-speed solar wind events. Burston et al. (2009) developed a physically based model to calculate the gradient drift instability growth on TEC patches monitored by GPS tomographic imaging, and they demonstrated a weak but significant linear correlation between GDI (Gradient Drift Instability) wave amplitude and the associated amplitude ($S_4$) and phase scintillation ($\sigma_\varphi$) indices. The best correlation was found with $\sigma_\varphi$. Burston et al. (2010) investigated the possibility for a turbulence process as an alternative to GDI, driven by short time fluctuations in the magnetospheric electric field, as originally proposed by Huba et al. (1985) and Kintner & Seyler (1985). They concluded that the GDI is most important but that the turbulent process might dominate at times. Carlson (2012) pointed out that the morphological picture currently in use for polar cap patches, on which the GDI rules all the plasma structuring, is an oversimplified picture that does not represent the current knowledge of the cusp/polar cap physics. He pointed out the need to differentiate between high-density polar cap patches (solar-EUV source) and low-density patches (produced by auroral particle precipitation). Strong flow shears giving rise to shear-driven instability must be taken into account in future research and development of mitigation techniques.

The cusp polar cap ionosphere is strongly influenced by the solar wind IMF conditions and the primary energy transfer mechanism from the solar wind into the magnetosphere-ionosphere system is believed to be impulsive dayside reconnection and Flux Transfer Events (FTEs) (Haerendel et al. 1978; Russell & Elphic 1978, 1979; Cowley & Lockwood 1992; Denig et al. 1993; Lockwood et al. 1995; Milan et al. 2000). FTEs are associated with flow channels (Van Eyken et al. 1984; Goertz et al. 1985; Lockwood et al. 1993; Pinnock et al. 1993, 1995; Moen et al. 1995; Provan et al. 1998, 2002; Neudegg et al. 1999, 2000; Provan & Yeoman 1999; Chisham et al. 2000; McWilliams et al. 2000; Oksavik et al. 2004a, Rinne et al. 2007, 2005), their associated current sheets (Taguchi et al. 1993; Moen et al. 2006, 2008b), and the formation of polar cap patches (Lockwood & Carlson 1992; Carlson et al. 2002, 2004, 2006; Lockwood et al. 2005; Moen et al. 2006; Lorentzen et al. 2010; Zhang et al. 2011) are of particular interest to ionospheric space weather physics research. The flow channel events call attention for the Kelvin-Helmholtz instability (KHI) to be considered as well.

Basu et al. (1988b, 1990) established a test based on simultaneous electron density and electric field spectra derived from Dynamics Explorer 2 (DE 2) data to discriminate between these two generic classes of instabilities operating in the high-latitude ionosphere. However, work still remains to be done to quantify the instability growth rates under various conditions. Moen et al. (2002) made an attempt to estimate growth rates for GDI based on electron density gradients derived from ionospheric tomographic images and plasma flow derived from SuperDARN measurements, but the growth rate they found was too slow to explain the close collocation of the equatorward cusp auroral boundary and the equatorward cusp HF radar backscatter boundary. (Hosokawa et al. 2009b) studied 630 nm airglow from polar cap patches and found that decameter scale irregularities observed by a SuperDARN radar extended over the entire region, which implies other structuring processes in addition to GDI (which requires a density gradient that is only found near the trailing edges of a patch). Carlson et al. (2007) reported evidence for a rapid development of plasma irregularities observed by scintillation (~10 s km–100 s m) in connection with patch formation above Svalbard.

They concluded that these irregularities had too rapid onset to be explained by any mechanism other than Kelvin-Helmholtz instability (KHI). They suggested a two-step mechanism where the KHI forms seed irregularities on which the GDI can operate. Since the FTE process is associated with both the formation of polar cap patches and flow channels, the hybrid GDI-KHI may be important in the cusp ionosphere, but has yet to be experimentally verified and quantified.

The major part of the current paper is devoted to review research under the COST action ES0803 to advance our understanding of the key processes giving rise to scintillation of radio waves in the polar cap ionosphere. In our investigations we apply coherent HF radar backscatter as a marker for plasma instability regions (Rodger et al. 1995; Yeoman et al. 1997; Milan et al. 1998, 1999; Moen et al. 2001, 2002; Oksavik et al. 2004b). The paper is organized as follows. Section 2 presents in-situ measurements of HF backscatter targets and a quantification of GDI growth rates in the cusp ionosphere. Section 3 presents reversed flow events (RFE) and a quantification of the KHI growth rate in connection with these events. Section 4 is devoted to a sequence of cusp flow channels with alternating IMF $B_Y$ giving rise to a sequence of flow channel events with alternating east-west flow direction resulting in strong flow shears, similar to the RFE category in Section 3. Section 5 describes a case study on tracing polar cap patches by the SuperDARN measurements and convection model. Section 6 shows how we can extract crucial information about plasma structuring from GPS scintillations climatology maps for Northern Norway and Svalbard. Section 7 summarizes the main findings.

## 2. GDI growth rates in the cusp ionosphere

Moen et al. (2002) proposed that the initial source of the decameter scale features responsible for the backscatter may result from fine structure within the precipitation itself, or alternatively, from cascade of unstable intermediate-scale gradients not resolved by the radio tomography technique. However, they pointed out an ultimate need for high-resolution in-situ measurements in order to assess the role of the GDI process in production of ionospheric irregularities, since tomographic imaging results are a coarse measure of electron density gradients.

During the COST action ES0803 we have investigated the GDI process by the Investigation of Cusp Irregularities sounding rocket mission (ICI-2). For this rocket mission the University of Oslo invented a new Langmuir probe experiment to provide absolute electron density measurements at several thousand kHz corresponding to submeter resolution for a rocket flight speed of 1 km/s. The Langmuir probe experiment consists of four identical cylindrical probes with a diameter of 0.5 mm and a length of 25 mm, and it is referred to as the 4-Needle Langmuir Probe system (4-NLP). On ICI-2 the four probes were biased at 2.5 V, 4.0 V, 5.5 V, and 10 V with respect to the rocket potential to measure the saturated electron currents (Bekkeng et al. 2010). A crucial feature of the 4-NLP measurement technique is that electron density measurement is insensitive to the electron temperature and the spacecraft potential (Jacobsen et al. 2010). The novel 4-NLP system can provide absolute electron densities of unprecedented resolution up to 10 kHz. For ICI-2 the 4-NLP sampling rate was 5.7 kHz which translates to submeter scale resolution during the flight.

The ICI-2 main objective was to perform in-situ observations of the coherent HF radar echoing targets and to quantify the gradient drift instability. Figure 1a shows that the rocket trajectory encountered the cusp ionosphere inflow region. Figure 1b, which actually displays 2 s of electron density measurements, demonstrates that we successfully resolved decameter scale HF radar backscatter targets. Moen et al. (2012) estimated the growth time for the GDI process to be 47.6 s in the direction along the rocket track. However, they pointed out that there was a significant plasma flow in the direction perpendicular to the rocket track. If a similar density gradient were to exist perpendicular to the direction of maximum plasma flow, the growth time would reduce to only 10 s. This is more than one order of magnitude faster than the ~10 min. growth time estimated by Moen et al. (2002). Oksavik et al. (2012) investigated spatial structures in the ICI-2 data, and Figure 2 shows a power spectral density plot of the electron density for the entire flight. Please note that the periods of 10-m scale irregularities extend from larger-in-amplitude irregularities at kilometer scale. It supports our assumption that regions of HF-backscatter are relevant to studies of GPS scintillations. The ICI-2 rocket also measured the electron precipitation, and the kilometer scale gradients lined up with beams of electron precipitation. Moen et al. (2012) concluded that kilometer scale electron density structures were created by ongoing soft precipitation on which the 10-m scale irregularities were generated at kilometer scale density gradients, possibly by the GDI process.

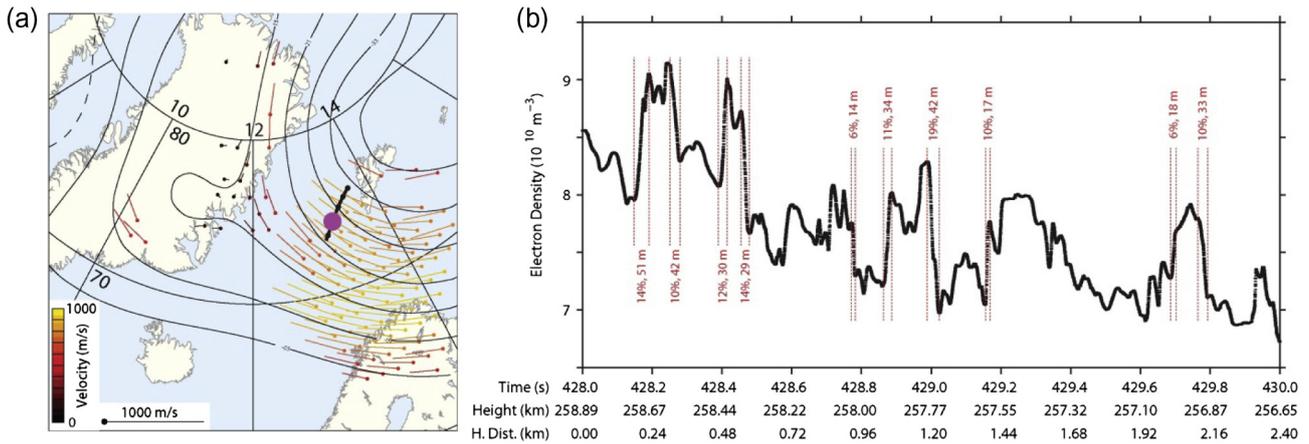

**Fig. 1.** (a) SuperDARN convection map with the ICI-2 flight overlaid. The coordinates are magnetic latitude and magnetic local time (MLAT, MLT). The black curve southwest of Svalbard represents the rocket trajectory above 200 km altitude. The rocket position at 10:42 UT (410 s flight time) is marked by a pink dot on the trajectory. (b) A zoom-in of 2 s high-resolution electron density data, where we have marked decameter scale plasma density gradients in red color. The altitude of the rocket payload, including horizontal distance traveled after 428 s, is shown as text along the time axis. The figure is adopted from Moen et al. (2012).

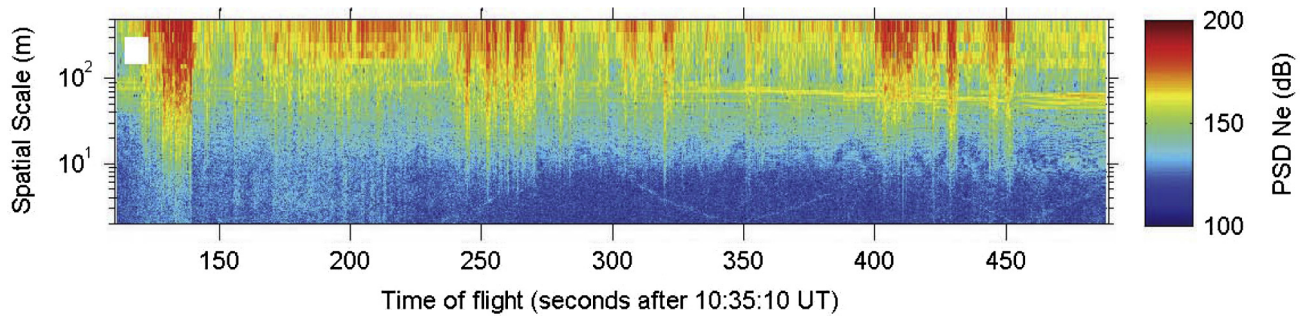

**Fig. 2.** Power spectrum of the electron density data during the ICI-2 flight, presented as spatial scales versus time of flight. The rocket payload is traversing a medium of variable plasma flow, so the horizontal separation $\Delta x$ between two data samples in the plasma frame of reference is $\Delta x = \Delta t\,[V_r + V_{E \times B}]$, where $\Delta t$ is the time between samples, $V_r$ is the horizontal speed of the rocket payload (around 1 km/s), and $V_{E \times B}$ is the component of $E \times B$ plasma drift along the rocket track. This technique has been used to obtain the spectral analysis of the spatial structures in the electron density. This figure demonstrates that 10-m scale structures extend from more intense irregularities at kilometer scale. The figure is adopted from Oksavik et al. (2012).

## 3. KHI growth rates associated with reversed flow events

Quite sophisticated radar modes have been developed to observe plasma flow dynamics (e.g., Carlson et al. 2002; Oksavik et al. 2004a, 2005, 2006). Employing the Special Norwegian Fast Azimuth Scan Mode (SP-NO-FASM) program at the EISCAT Svalbard Radar (ESR), Rinne et al. (2007) discovered a new category of flow channels, Reversed Flow Events (RFEs), reversed by means of flow opposing the background convection. Figure 3a shows an example of an RFE event in a 120° wide ESR scan from (180° to 300° azimuth measured clockwise from 0° at geographic north) at 30° elevation. The background plasma convection monitored by the ESR is uniform and predominantly directed westward and away from the radar (red color), consistent with the SuperDARN convection map in Figure 3b, except from the southern part of the scan where the ESR line-of-sight velocity is not sensitive to zonal flow. The RFE channel is the blue (eastward) channel opposing the red background (westward). RFEs are ~100–200 km wide east-west elongated channels that have an average lifetime of ~18 min, i.e., quasi-static flow structures. In order to qualify as an event, the line-of-sight velocity inside the RFE must be larger than 250 ms$^{-1}$ and oppose the background flow in at least one scan. Rinne et al. (2007) found RFE occurrence in 40% of their ESR scans in the 1140–1250 MLT sector, with no apparent preference for IMF $B_Y$ or $B_Z$ polarity, but a clear preference for clock angles between 40° and 240°. The RFE is therefore a cusp phenomenon that deserves particular attention in polar cap space weather research.

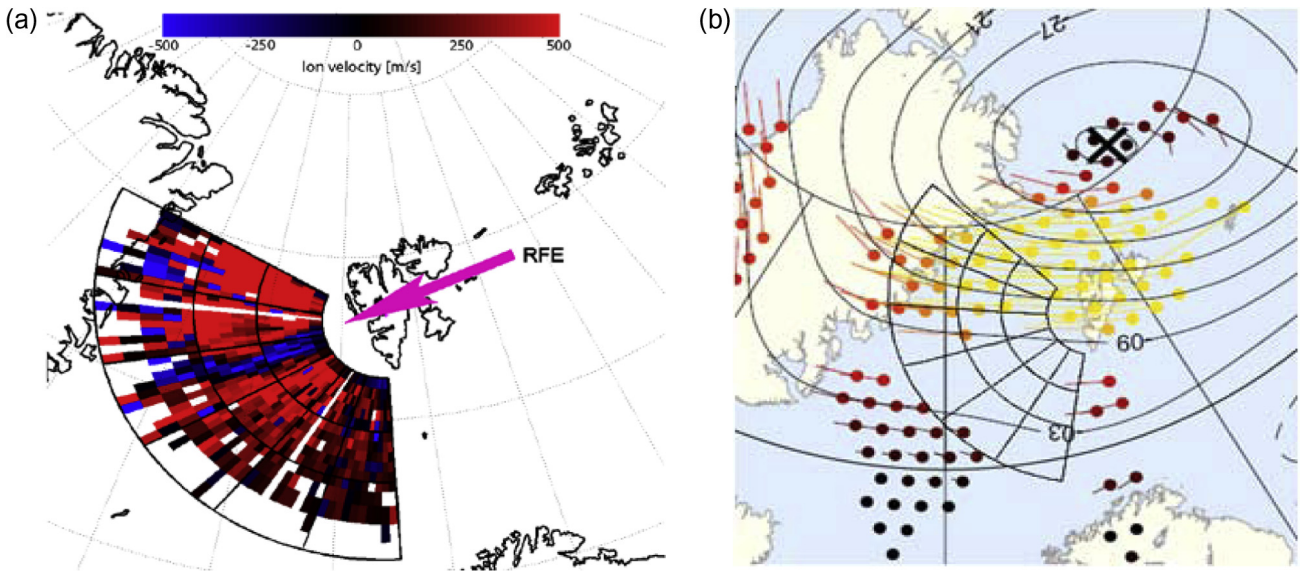

**Fig. 3.** (a) An example of an RFE event of eastward flow (blue) opposing the westward background flow (blue). (b) SuperDARN convection map with westward convection over Svalbard (directed toward magnetic noon).

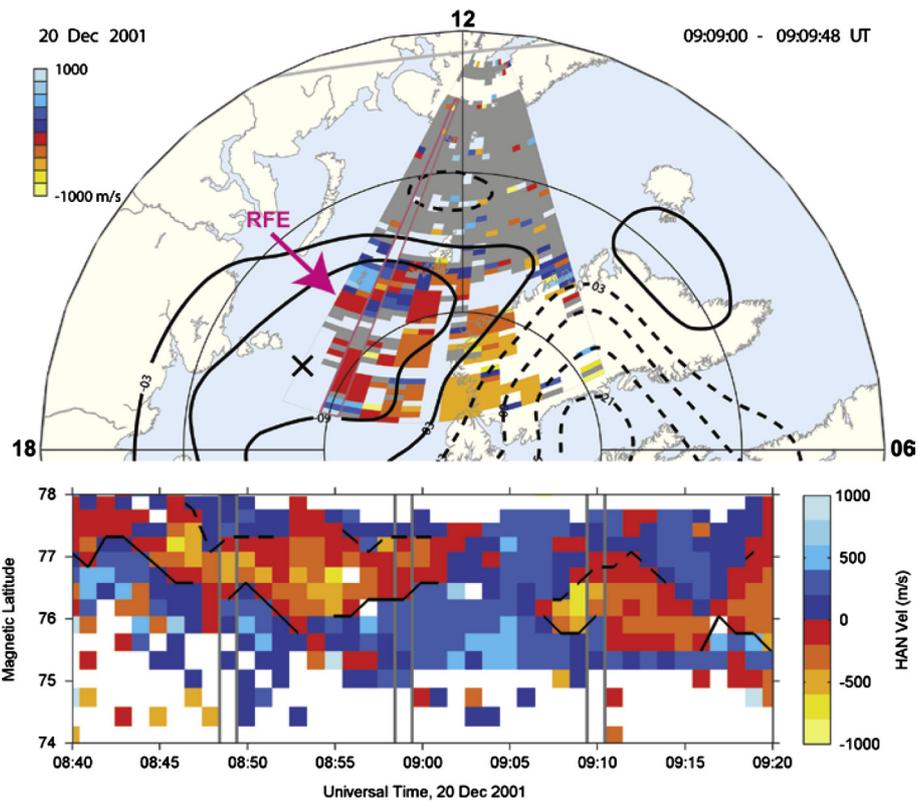

**Fig. 4.** The top panel shows an RFE channel on the dayside seen by SuperDARN, i.e., the narrow channel of antisunward flow (red color) embedded in a background of sunward flow (blue color). The lower panel shows the time evolution along one of the SuperDARN beams. The figure is adopted from Oksavik et al. (2011).

According to Moen et al. (2008b) the RFE phenomenon does not appear to be uniquely related to PMAFs. It rather appears to be a specific feature of Birkeland Current Arcs (BCAs). Moen et al. (2008b) provide two possible explanations for the generation of RFEs: (1) the RFE channel may be a region where two MI current loops, forced by independent voltage generators, couple through a poorly conducting ionosphere, or (2) the RFE channel may be the ionospheric footprint of an inverted-V-type coupling region. It is still unclear if one of the two mechanisms dominates, or if both mechanisms are closely related.

During COST action ES0803 Oksavik et al. (2011) reported the first clear case of RFE channels seen in SuperDARN HF radar data, presented in Figure 4. The flow shears of RFEs are associated with wide Doppler spectra and may cause a rapid development of shear-driven instabilities. In their scan data Oksavik et al. (2011) found several examples of an immediate response in enhanced backscatter power, which implies rapid development of decameter irregularities within the 1-min resolution of the radar scan. Oksavik et al. (2011) estimated a KH growth time of the order of 1–3 min using line-of-sight velocity data from the SuperDARN Hankasalmi radar. This result substantiates the Carlson et al. (2007) two-step KHI-GDI mechanism, but the problem is that we have no density data. The ICI-3 rocket was launched on 3 December 2011 to investigate closer the RFE instability processes, and this issue is under investigation. The huge SuperDARN data set (more than one decade) opens a new opportunity for systematic studies of the large-scale characteristics of RFEs and flow channels, including how plasma instabilities operate in the cusp region more generally.

## 4. Stacked cusp flow channel events

Rinne et al. (2010, 2011) investigated a sequence of stacked flow channel events (FCEs) where the zonal flow direction alternated from east to west in response to IMF By polarity changes. On the background of northwestward convection in the postnoon sector, consistent with prevailing IMF $B_Y$ positive and IMF $B_Z$ negative, a sequence of three eastward flow channels formed in response to three sharp IMF rotations to $B_Y$ negative and $B_Z$ positive. Figure 5 shows line-of-sight ion velocities obtained during the 11:04:59–11:08:07 UT ESR scan on 06 September 2005. The color scale is the same as in Figure 3a, ranging from −500 m/s (blue toward) to 500 m/s (red away). The two blue eastward flowing channels E1 and E2 in Figure 5, interspaced by a westward flow channel (W1), are interpreted as the footprint of a sequence of three subsequent flux transfer events. E1 and W1 formed at the same location as E2 in the present image and propagated poleward in a tandem motion. The alterations in east-west flow direction were found to be consistent with alterations in the magnetic tension force due to IMF $B_Y$ polarity changes. The observations are consistent with the view that a new region of reconnected flux develops as a distinct flow channel near the polar cap boundary, and successive events remain separated while pushing each other into the polar cap (Lockwood et al. 2001). Each flow channel will remain separated from neighboring channels mapping to different reconnection sites as long as the magnetic tension force with its associated field aligned current systems is maintained. E2 looks like an RFE channel, but it is not, since the direction of flow is consistent with the IMF $B_Y$ magnetic tension force. However, regarding the GDI instability the eastward flow channels in Figure 5 are of the same caliber as the RFE channels in Section 3.

Figure 6 presents concurrent observations of the development of the first eastward channel, E1, between 10:45 and 10:59 UT by ESR and The Cooperative UK Twin Auroral

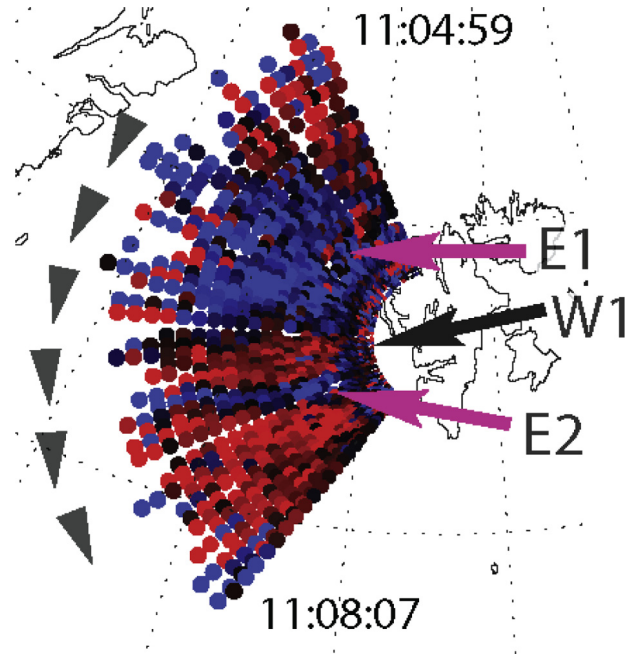

**Fig. 5.** ESR line-of-sight velocities for the radar scan at 11:04:59–11:08:07 UT on 06 September 2005, showing three distinct flow channels in the cusp region. Two eastward flow channels (E1 and E2 in blue) are interspaced by a westward flow channel (W1 in red). The color scale is the same as in Figure 3a.

Sounding System (CUTLASS) Pykkvibaer radar in Iceland. Comparing the ESR (left column) and Pykkvibaer radar ($v_{los}$ middle column, and backscatter power right column), the region of HF backscatter apparently developed in phase with the development of the E1 flow channel, but with a small time lag. The initial phase of E1 was detected by the ESR at 10:43 UT, and the blue flow channel was well established one scan later at 10:48 UT. The CUTLASS radar detected the development of HF backscatter with the initialization of E1 (cf. Fig. 6, right column), which indicates that backscatter irregularities formed within the 2 min cadence time of the radar observations. This indicates that efficient instability processes are associated with FTE flow channels.

## 5. Polar cap patch formation and propagation

The solar-EUV ionized plasma enters the polar cap through the cusp inflow region and forms the tongue of ionization (TOI) (Knudsen 1974; Foster et al. 2005; Moen et al. 2008a). Sato (1959) was the first to postulate the existence of this TOI, which was initially regarded as a homogeneous stream of high-density plasma from a sunlit dayside source region across the polar cap to the dark nightside by the solar wind driven twin-cell E × B drift (Buchau et al. 1983, 1985; Foster & Doupnik 1984; Weber et al. 1984). The TOI can either be a continuous region, or it can be sliced into a series of discrete substructures (Moen et al. 2006; Hosokawa et al. 2009b), where flux transfer events appear to be an efficient cutting mechanism (cf. Sect. 1).

Oksavik et al. (2010) combined observations from ESR and SuperDARN to study events of extreme plasma density ($n_e \geq 10^{12}$ m$^{-3}$) that drifted across the polar cap, as demonstrated in Figure 7. The events occurred in a period when IMF $B_Y$ was strongly positive. The ESR first detected the events near the cusp inflow region, and the wide SuperDARN data coverage was used to follow the plasma motion. Oksavik et al. (2010)

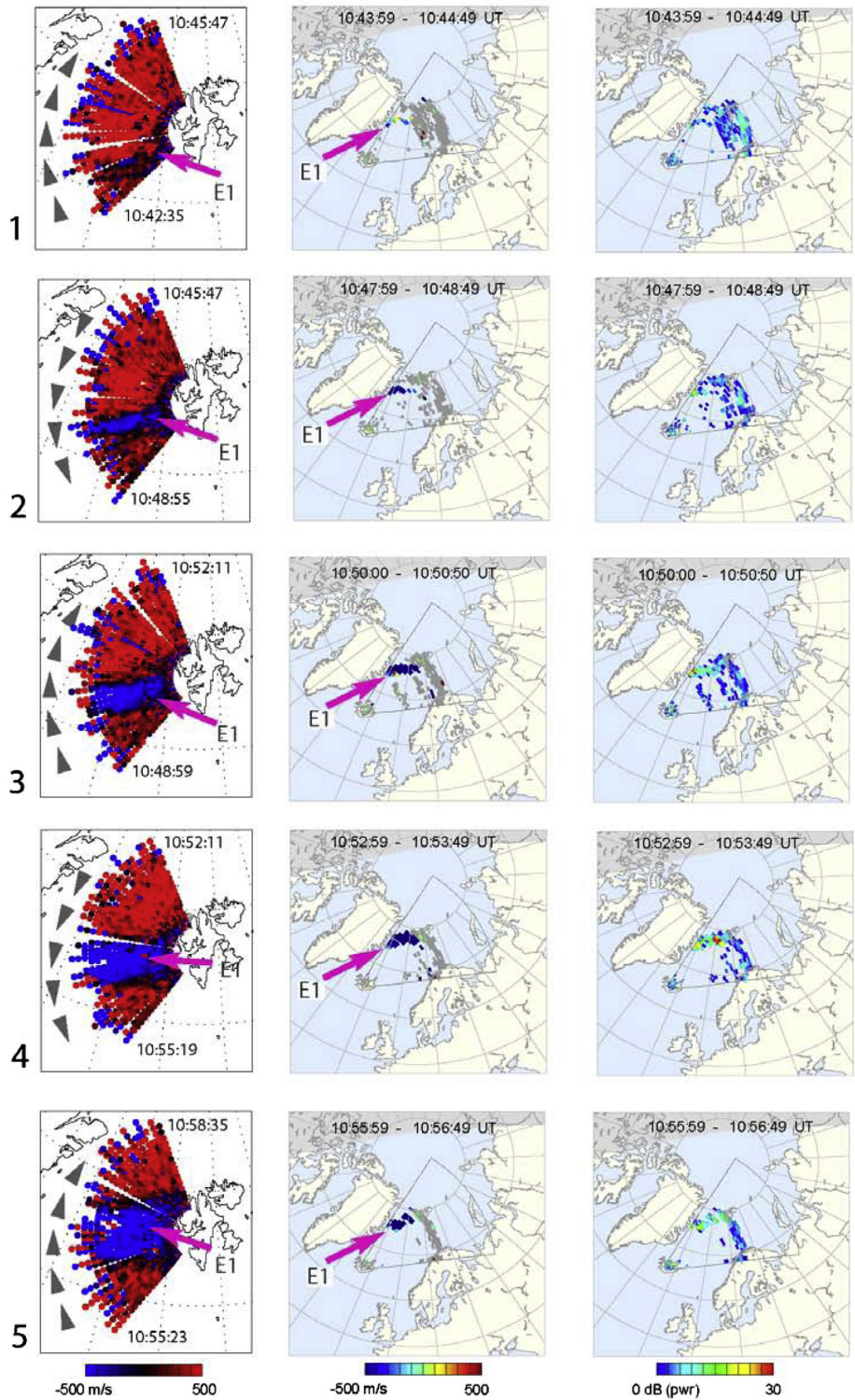

**Fig. 6.** Line-of-sight velocity data from five consecutive ESR scans from 10:45:47–10:58:35 UT on 06 September 2005 (left column). The 32 m steerable ESR antenna swept back and forth in azimuth at a fixed elevation angle of 30° in a windshield wiper mode at a cadence of 192 s. Corresponding line-of-sight velocity data from the Pykkvibaer HF radar in Iceland (middle column), and backscatter power data from the Pykkvibaer HF radar (right column).

introduced a tracking algorithm that applies the traditional SuperDARN convection patterns (Ruohoniemi & Greenwald 1996; Ruohoniemi & Baker 1998) to obtain an array of flow vectors across the entire polar cap at 2-min cadence and with a grid resolution of 1° MLAT and 2° MLON. Then they used these SuperDARN convection data to trace the extreme densities forward and backward in time by injecting a test particle every minute at the ESR location (i.e., 75.37° MLAT and 113.94° MLON). The MLT of each data point was determined, and the closest flow vector (maximum 1° MLAT or 0–115 km away) was used to trace the test particle toward the next minute boundary. This new location was then used as the next starting point, and the process was repeated. In total each test particle was traced 2 h backward and 2 h forward in time. By this new method they demonstrated that events originated in the afternoon sector at lower sunlit latitudes, before the large-scale convection transported the dense plasma into and across the polar cap. The SuperDARN Hankasalmi radar in Finland confirmed how one of the events actually rotated in a clockwise direction once it entered the dayside polar cap. The SuperDARN Kodiak radar in Alaska verified the subsequent time of arrival on the nightside (i.e., confirming the accuracy of the tracing across the polar cap from Europe to Alaska). During the transit across the polar cap the patch propagation speed was both pulsed, and the plasma patch underwent substantial rotation. A similar deformation of polar cap patches in the central polar cap has also been confirmed very recently by Dahlgren et al. (2012a, 2012b) using the Resolute Bay Incoherent Scatter Radar (RISR-N) in northern Canada.

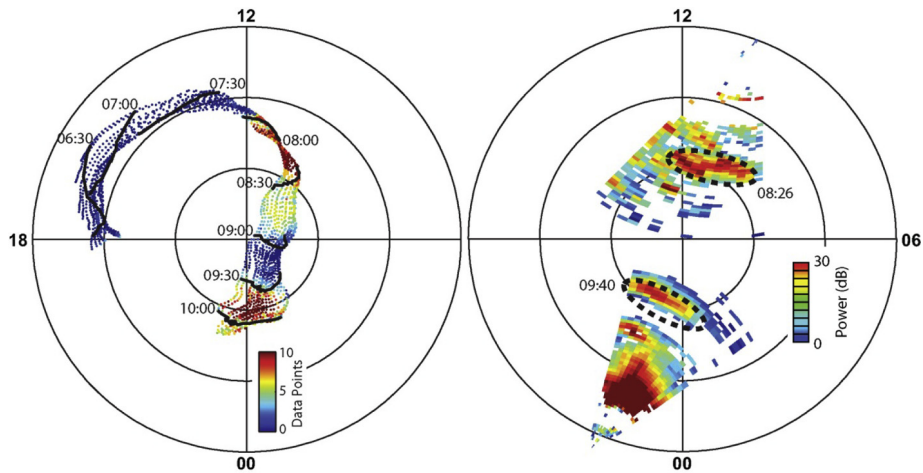

**Fig. 7.** The left panel shows the propagation across the polar cap of an area of extreme plasma density. The right panel shows that one SuperDARN radar on the dayside (at 0826 UT) and another SuperDARN radar on the nightside (at 0940 UT) detected enhanced backscatter power when the same event passed by. The figure is adopted from Oksavik et al. (2010).

The study of Oksavik et al. (2010) highlights several new findings: in the past, attention has mainly been on what happens to polar cap patches when they are created on the dayside near the open/closed field line boundary. Less attention has been devoted to the detailed propagation pattern while in transit across the polar cap, implicitly assuming that they just follow the antisunward flow. The study of Oksavik et al. (2010) shows that this simplified picture must be advanced. Momentum transfer can last significantly longer than 10 min after reconnection, especially for extremely long field lines where IMF $B_Y$ is dominating, i.e., on "old open field lines". The flow speed across the polar cap can also be highly dynamic and pulsed. Patches can undergo substantial rotation in the central polar cap, so that the leading edge can become the trailing edge, and the first patch to enter the dayside polar cap may not be the first patch to reach the nightside. Plasma can be stagnant in the polar cap, or even overtaken. The strong gradients in the plasma flow associated with the rotation may in itself also further enhance the growth of ionospheric irregularities all the way across the polar cap. With a quickly growing chain of new SuperDARN radars from mid-latitudes to the central polar cap, and the new capability of obtaining three-dimensional volumetric imaging of polar ionospheric structures over Resolute Bay (e.g., Dahlgren et al. 2012a, 2012b), there are unprecedented opportunities in the years ahead to further explore this phenomenon, and possibly develop a tool that can monitor the propagation of these events in real time.

The fact that the patch light up in the SuperDARN radar field-of-view confirms that there are ongoing instability processes and generation of HF backscatter targets. However, the solar wind-IMF-driven dynamics of plasma intake, and the non-constant flow pattern during their transit across the polar cap, poses challenges when it comes to developing forecast models. However, the tracking algorithm developed by Oksavik et al. (2010) provides a new tool.

## 6. Scintillation climatology studies

A number of networks located at auroral and polar cap latitudes provide a good statistics of GNSS measurements acquired at 50 Hz frequency sampling (Alfonsi et al. 2011; Prikryl et al. 2011a, 2011b). The information derived from the data can provide clues to understand the cause-effect mechanisms producing ionospheric scintillations. Such archives enable scintillation climatology; an assessment of the recurrent features of the ionospheric irregularity dynamics and temporal evolution on long data series, with the attempt to catch possible correspondences with scintillation occurrence.

There is currently no available technique to resolve temporal and spatial variability of the F-region plasma irregularities and some assumptions have to be made in order to further quantify the scintillation source targets. For the purpose of assessing the spatial structure of scintillation irregularities, Alfonsi et al. (2011) attempted to catalog the occurrence of phase and amplitude scintillations according to the mean and sigma values of the Rate of TEC changes (ROTs). Their approach is based on the following assumptions. ROT is the time rate of change of the differential carrier phases, providing complementary information about which electron density irregularity scale sizes give rise to GNSS scintillations. ROT is computed over 1 min intervals, resulting in a Nyquist period of 2 min. The corresponding scale length is given by the components of the ionospheric projection of the satellite motion and the irregularities in a direction perpendicular to the propagation path. Assuming that the plasma convection velocity at high latitudes is between 100 m/s and 1 km/s, the irregularity scale lengths sampled by ROT span from a few kilometers to tens of kilometers (Basu et al. 1999). The amplitude scintillation, differently from the phase scintillation, is biased by irregularities probing size, which on $L$ band is hundreds of meters (Aarons 1997, and references therein). Under these assumptions

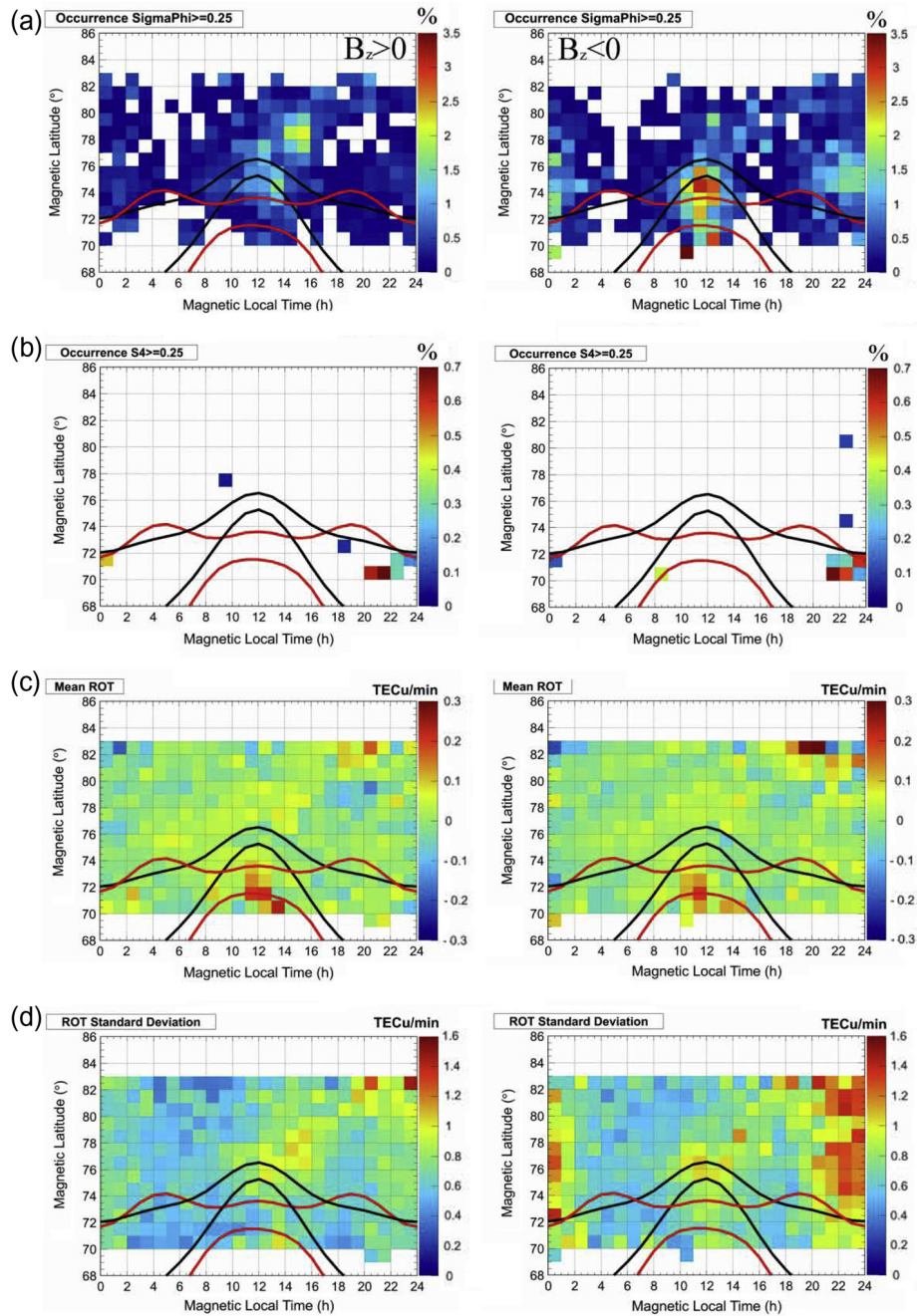

**Fig. 8.** Ionospheric climatology observed between October and December 2003 over Ny-Ålesund obtained by binning the data for conditions of IMF $B_Z > 0$ (left column) and $B_Z < 0$ (right column). Black and red curves reproduce the Feldstein model auroral ovals for IQ = 3 and IQ = 6, respectively. From the top: (a) percentage occurrence of phase scintillation, and (b) percentage occurrence of amplitude scintillation, (c) ROT, and (d) standard deviation of ROT.

Alfonsi et al. (2011) addressed the amplitude scintillations (S4 index) to scale sizes up to hundreds of meters, ROT to larger scale sizes (few kilometers), and phase scintillations to irregularities of all scale sizes.

Figure 8 shows the ionospheric climatology as observed from Ny-Ålesund in Svalbard (78.9°N, 11.9°E; CGM 76.0°N, 112.3°E) between October and December 2003. The plots, obtained by means of the Ground-Based Scintillation Climatology (GBSC) method introduced by Spogli et al. (2009), have been sorted according to IMF direction by selecting $B_Z > 0$ (left column) and $B_Z < 0$ (right column) conditions, analogously to the study of Alfonsi et al. (2011). The attempt is to investigate the role of the IMF $B_Z$ ionosphere interplay in driving the scintillation phenomena at high latitude. The period was characterized by two great storms and notable scintillation events on 30–31 October and on 20–21 November 2003 (e.g., De Franceschi et al. 2008). The black and red curves in Figure 8 reproduce the Feldstein model auroral ovals for IQ = 3 (disturbed) and IQ = 6 (highly disturbed) conditions, respectively. The climatological picture reveals, as reported by Spogli et al. (2009), that even under extremely disturbed conditions the polar cap ionosphere is not affected, except over a few isolated sectors, by moderate/strong amplitude scintillation (Fig. 8b), independent of the IMF orientation. It indicates that, on average, the fragmentation of the tongue of ionization around noon does not result in amplitude scintillations exceeding the $S_4 > 0.25$ threshold. The only sites where amplitude scintillation has a small chance to occur are the boundaries of the auroral oval, and in particular the pre-midnight MLT sector.

Concerning phase scintillation (Fig. 8a), under southward IMF conditions magnetopause reconnection leads to soft particle precipitation in the cusp region and is associated with larger occurrence of phase scintillation than under northward IMF conditions. It is interesting to note how the cusp footprint is reproduced in the phase scintillation occurrence; the spread and the equatorward shift of the cusp are evident in the phase scintillation occurrence when $B_Z$ is negative, while the cusp signature occurs at higher latitudes when $B_Z$ is northward. Since TOI patches are not expected for northward IMF, the presence of cusp scintillations for northward IMF indicates that particle precipitation plays a central role in irregularity formation.

By looking at differences between $B_Z$ positive and negative in the |ROT| and ROT standard deviation (ROT_SD) maps (Fig. 8c and 8d), the general behavior of |ROT| is not significantly different between the two IMF orientations; in both maps |ROT| enhances in the cusp region. ROT_SD maps are quite different, highlighting larger values in the polar cap at postnoon (about 1 TECu/min between 14 and 18 MLT) under $B_Z > 0$. It indicates that, during northward IMF orientation, inside the polar cap there are irregularities on a wide range of different scale sizes. When $B_Z$ is negative, larger values (1.4–1.6 TECu/min) of ROT_SD are spread poleward over a wider range of MLATs, still in the polar cap, around midnight. This could be explained by the presence of patches in the polar cap that are more likely to occur under such IMF orientation (Alfonsi et al. 2011, and references therein). The asymmetry in ROT_SD, which favors the pre-midnight magnetic sector, is another signature of the presence of patches (Moen et al. 2007). Considering the catalog proposed in Table 3 of the paper by Alfonsi et al. (2011), a general agreement is found between the expected relationships of |ROT|, ROT_SD, irregularities in scale sizes, and scintillation occurrence. In fact, larger values of ROT_SD (Fig. 8d) are always associated with the presence of several scale sizes, possibly leading to both amplitude and phase scintillations (Fig. 8a and 8b). According to the same catalog, the larger values of ROT_SD (Fig. 8d) when $B_Z < 0$ lead to the enhancement of phase scintillation possibly due to patches in the pre-midnight sector.

## 7. Summary and concluding remarks

During southward IMF trains of polar cap patches/solar-EUV ionized plasma drift across the polar cap from day to night (Foster 1984; McEwen & Harris 1995, 1996; McEwen et al. 1995, 2004; Lorentzen et al. 2004; Foster et al. 2005; MacDougall & Jayachandran 2007; Moen et al. 2007, 2008a; Hosokawa et al. 2006, 2009a, 2011). While the flow channels are more localized, the polar cap patches drift across the polar cap from day to night and may be active several hours as a source for radio wave scintillations. Scintillation of trans-ionosphere radio signals in the polar cap is a space weather issue of increasing interest as offshore activities move to higher latitudes, with high precision needs for satellite navigation, and there are growing interests to use the Northern Ocean Route between Europe and Asia, while the aviation industry exploits more extensive use of GNSS services. It is well known that these activities are connected to the energy transfer from the solar wind to the ionosphere, which is related to the solar and geomagnetic activities. Theory and modeling work of plasma instability processes and turbulence has been reported in the literature. However, these are yet to be verified and quantified in order to yield building blocks for future scintillation forecast models which is an ultimate space weather product.

During the COST action ES0803 we have conducted several in-depth studies of the underlying physics of plasma instabilities related to flow channels near the polar cap boundary and the study of polar cap patches and density structures. Scintillation climatology studies have been conducted in order to identify the regions where the largest scintillation disturbances are expected. The COST action ES0803 has been carried during an extended period of low solar activity and the large fraction of the work reviewed here is based on data from the years 2001–2003 during the declining phase of solar cycle 23.

It is evident from the scintillation maps that the polar cap ionosphere is subject to phase scintillations but not to amplitude scintillations. The largest scintillation disturbances occur during intervals of IMF $B_Z$ negative conditions and are located in the cusp inflow region around local magnetic noon (10–14 MLT) and near the exit region 20–02 MLT around magnetic midnight, concurrent with formation/intake and exit of polar cap patches, consistent with Spogli et al. (2009) and Prikryl et al. (2010). However, our studies show that further refinement is needed to include the effects of particle impact ionization and flow shears.

The studies by radar and rocket measurements have during this COST action focused on the cusp region, where Svalbard is ideally located and equipped with both incoherent scatter radar (ESR) and a launch site for sounding rockets. With the unprecedented resolution of the ICI-2 rocket data we were able to quantify the growth rate of the gradient drift instability process related to kilometer scale plasma gradients produced/enhanced by cusp electron precipitation. The growth rate was of the order of tens of seconds, which means that this GDI process may efficiently spawn structures from kilometer scale structures formed by cusp auroral filaments. This further underscores the view that scintillation irregularities do not exclusively grow on large-scale density gradients of polar cap patches. The elevated ROT_SD values and the phase scintillations enhancement in the cusp region are consistent with the multi-scale structures observed by ICI-2.

Using both the ESR and SuperDARN we have distinguished two subcategories of 100–200 km wide flow channels, the RFEs (reversed flow opposing the background flow, i.e., the flow direction is not consistent with the IMF $B_Y$ magnetic tension force) and the FCEs (newly reconnected flow channels where the flow direction is consistent with the IMF $B_Y$ magnetic tension force of newly open flux). The Kelvin-Helmholtz instability may efficiently operate near the flow shear for both categories. We have found evidence for the two-step KHI-GDI process suggested by Carlson et al. (2007). In December 2011 we intersected the RFE flow channel with the ICI-3 sounding rocket, but data is still under investigation. Hopefully, we will be in a position to quantify the growth rate of the KHI operating in the cusp inflow region.

We have presented a promising technique employing SuperDARN convection maps to track patches detected by the EISCAT Svalbard Radar both forward and backward in time. The pulsed flow dynamics will be a challenge to precisely forecast the transit time across the polar cap and the arrival of polar cap patches at night, but quite accurate 1–2 h forecasts should be possible. The ROT_SD maps indicate that the multi-scale structure of polar cap patches is maintained during their transit toward the midnight sector.

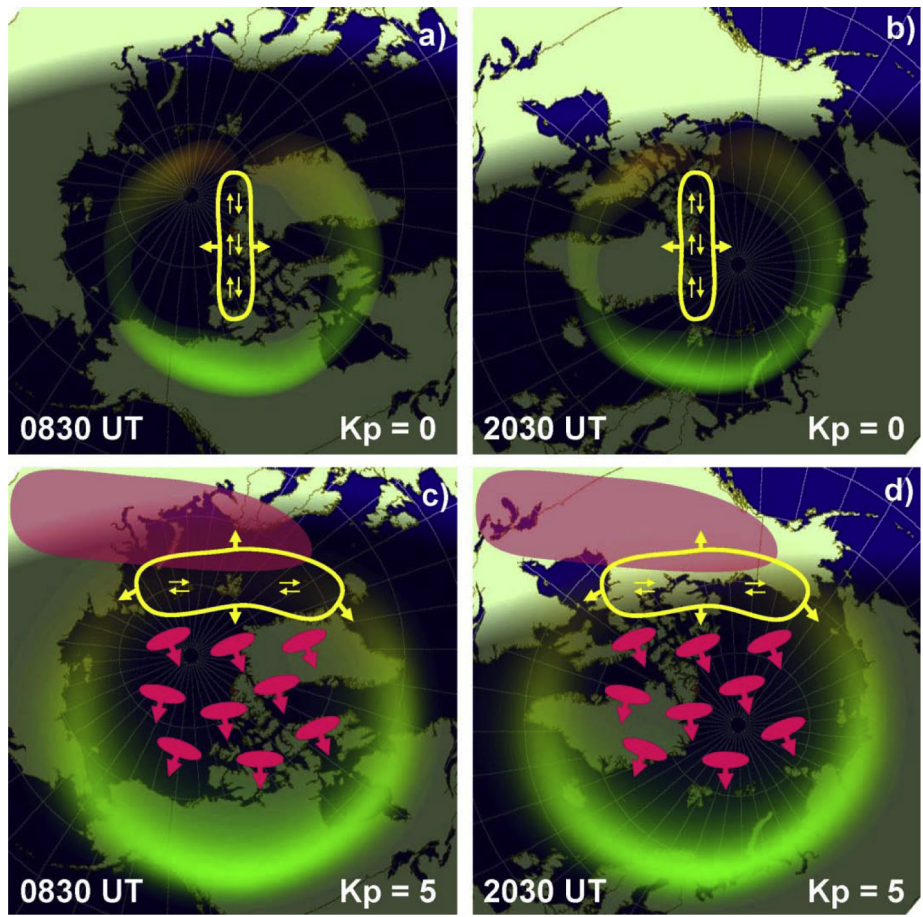

**Fig. 9.** A schematic illustration of active space weather regions in the polar cap ionosphere when IMF $B_Z$ is north (top row, $Kp = 0$), and IMF $B_Z$ is south (bottom row, $Kp = 5$). The active regions for creation of polar cap patches/plasma irregularities are shown in yellow color and move under the influence of IMF as indicated with yellow arrows. For IMF $B_Z$ north the active region is caused by flow shears near transpolar arcs in the central polar cap, and space weather problems are only expected far north of Svalbard both day (panel a, 0830 UT) and night (panel b, 2030 UT). For IMF $B_Z$ south the tongue of ionization (pink) extends into the dayside auroral oval, where magnetic reconnection chops it into polar cap patches (pink) that begin to drift across the polar cap. In the production region there are flow channels and strong flow shears that initiate the growth of ionospheric irregularities. Svalbard will be directly under the production region at daytime (panel c, 0830 UT), and at night Svalbard will see patches arriving from the polar cap (panel d, 2030 UT).

The two bottom panels of Figure 9 summarize the situation for active conditions for IMF $B_Z$ south. The auroral oval corresponds to $Kp = 5$. Panel c is at 08:30 UT when the active cusp is over Svalbard, and panel d is at 20:30 UT when Svalbard is near magnetic midnight. For IMF $B_Z$ south and dayside magnetopause reconnection the polar cap scintillation irregularities are featured by flow channels and formation of plasma gradients near the cusp inflow region, and by the transport of polar cap patches across the polar cap that exit the polar cap due to tail reconnection. For IMF $B_Z$ north, illustrated by the two top panels a and b ($Kp = 0$), in addition to auroral precipitation in the oval, we expect transpolar arcs and flow shears in the polar cap that potentially may generate irregularities (Carlson 2012). According to the climatology, patches are less of an issue for IMF $B_Z$ north. However, the presence of phase scintillation in the cusp for northward IMF indicates that cusp electron precipitation drives plasma instabilities, emphasizing the potential role of GDI on auroral plasma structuring, as revealed by the ICI-2 rocket study (Moen et al. 2012).

*Acknowledgements.* We thank the SuperDARN PIs, and in particular Prof. Mark Lester at the University of Leicester, for provision of the SuperDARN radar data. The ICI-2 rocket operation was supported by the Andøya Rocket Range. EISCAT is an international association supported by research organizations in Norway (NFR), Sweden (VR), Finland (SA), Japan (NIPR and STEL), China (CRIRP), the United Kingdom (STFC), Germany (DFG, till 2011), and France (CNRS, till 2005). This project has also been sponsored by the Research Council of Norway, the Air Force Office of Scientific Research, Air Force Material Command, USAF, under Grant No. FA8655-10-1-3003, Programma Nazionale di Ricerche in Antartide (PNRA), CNR (Consiglio Nazionale delle Ricerche), and COST action ES0803.